\documentclass[twocolumn,floatfix,prb,aps,showpacs,longbibliography]{revtex4}

\usepackage{graphicx}
\usepackage{xcolor}
\usepackage{bm}
\usepackage{amsmath}
\usepackage{amssymb}
\usepackage{amsthm}
\usepackage{amsfonts}
\usepackage[english]{babel}
\usepackage{float}
\usepackage{tipa}
\usepackage{mathtools}

\usepackage{bbm}

\begin{document}

\title{Elementary Excitations in Fractional Quantum Hall Effect from Classical Constraints}
\author{Bo Yang$^{1,2}$ and Ajit C. Balram$^{3,4}$} 
\affiliation{$^{1}$Division of Physics and Applied Physics, Nanyang Technological University, Singapore 637371.}
\affiliation{$^{2}$Institute of High Performance Computing, A*STAR, Singapore, 138632.}
\affiliation{$^{3}$The Institute of Mathematical Sciences, HBNI, CIT Campus, Chennai 600113, India}
\affiliation{$^{4}$Niels Bohr International Academy and the Center for Quantum Devices, Niels Bohr Institute, University of Copenhagen, 2100 Copenhagen, Denmark}


\date{\today}
\begin{abstract}
Classical constraints on the reduced density matrix of quantum fluids in a single Landau level termed as local exclusion conditions (LECs) [B. Yang, Phys. Rev. B {\bf 100} 241302 (2019)], have recently been shown to characterize the ground state of many fractional quantum Hall (FQH) phases. In this work, we extend the LEC construction to build the elementary excitations, namely quasiholes and quasielectrons, of these FQH phases. In particular, we elucidate the quasihole counting, categorize various types of quasielectrons, and construct their microscopic wave functions. Our extensive numerical calculations indicate that the undressed quasielectron excitations of the Laughlin state obtained from LECs are topologically equivalent to those obtained from the composite fermion theory. Intriguingly, the LEC construction unveils interesting connections between different FQH phases and offers a novel perspective on exotic states such as the Gaffnian and the Fibonacci state. 
\end{abstract}

\maketitle 

\section{Introduction}

Low lying elementary excitations in topologically ordered systems are fascinating objects that capture the essential topological features of the corresponding ground states. In fractional quantum Hall (FQH) systems, these excitations can carry fractionalized charges and obey anyonic or non-Abelian braiding statistics~\cite{Tsui82, Laughlin83, Arovas84, Moore91}. On one hand, discerning the nature of these excitations offers important insights into the underlying mechanisms of incompressibility of the FQH states. On the other hand, manipulating these excitations in a controlled manner also holds promises for storing and processing quantum information that is topologically protected~\cite{Kitaev03, Kitaev06, Nayak08, Freedman02}. Thus, a detailed understanding of the nature of the elementary excitations of FQH states is important from both the fundamental as well as the practical standpoint.

The elementary excitations in FQH systems can be categorized into charged and neutral excitations. The former includes the positively charged quasiholes and the negatively charged quasielectrons. The quasihole excitation can be realized by inserting flux quanta into the incompressible quantum fluid. For FQH states with model Hamiltonians~\cite{Laughlin83, Haldane83, Moore91, Read99, Simon07a}, the quasihole states usually form a zero energy manifold degenerate with the ground state. The counting of the quasihole states on the sphere encodes the topological order of the FQH phase, from which one can derive the edge modes on the disk or ground state degeneracy on the torus. Whether an FQH state is Abelian or non-Abelian is also determined by the quasihole properties. In particular, a state is non-Abelian if just specifying the positions of the quasiholes does not uniquely determine the wave function for the multi-quasihole state.

Following the terminology used in Ref.~\cite{Yang14}, we use the term ``quasielectrons", instead of ``quasiparticles", for elementary charged excitations carrying fractionalized negative charges in an FQH system. The quasielectrons can be created by adding electrons to or removing fluxes from the FQH fluid. A neutral excitation is composed of a quasielectron and a quasihole. The low-lying branch of the neutral excitations of FQH fluids forms a collective mode called the magnetoroton mode~\cite{Girvin85, Girvin86}. A given FQH phase can host different types of quasielectron and neutral excitations~\cite{Yang14, Majumder09, Balram16b}. For the FQH state to be incompressible, \emph{all} the quasielectrons, as well as the neutral excitations, have to be gapped in the thermodynamic limit.  

While the construction of the quasihole states is relatively straightforward, the construction of model states to represent quasielectrons and neutral excitations is much more involved. This is because, in contrast to the ground state and quasiholes, model Hamiltonians for quasielectrons and neutral excitations are not known. For many FQH states, the composite fermion (CF)~\cite{Jain89} theory or the Jack polynomial~\cite{Bernevig08} and conformal field theory (CFT)~\cite{Hansson17} approach can be used to construct excitations involving quasielectrons. These different approaches generically lead to different microscopic wave functions for the same topological state of quasielectrons~\cite{Bernevig09b}. It is also worth noting that while for each FQH phase there is only one type of quasiholes, there can be in general many different types of gapped quasielectrons and neutral excitations~\cite{Yang14, Majumder09, Balram16b}. For some exotic non-Abelian phases, these gapped excitations are not very well understood.

Previously, one of us introduced the idea of local exclusion conditions (LECs) in the FQH effect and demonstrated that the LECs in conjunction with the requirement of translational invariance can determine the topological properties of many FQH states~\cite{Yang19}. A LEC is a well-defined constraint on the reduced density matrix of a quantum Hall fluid. Physically it is a constraint on what we can or cannot measure, thus in a sense, it is a ``classical constraint" on a quantum system. Each LEC is specified by a triplet of integers $\hat n=\{n,n_e,n_h\}$. An imposition of $\hat n$ on a quantum Hall fluid dictates that for any circular droplet within the fluid containing $n$ fluxes, a physical measurement can neither detect more than $n_e$ number of electrons, nor $n_h$ number of holes. Topological indices including filling factor, shift~\cite{Wen92} and particle clustering~\cite{Read99} can emerge from quantum Hall fluids satisfying the LECs. Furthermore, LECs also determine the microscopic model wave function of the many-body ground state for these topological phases. One should also note that for many FQH states (Read-Rezayi sequence, Gaffnian, Haffnian, etc.) the number of translationally/rotationally invariant, i.e., $L=0$, states is exactly one in the truncated basis satisfying the LEC. Such states can plausibly lend themselves to a description in terms of a LEC. However, there are also states, like ground states of hollow-core Hamiltonians or the unprojected $2/5$ Jain CF state, for which this is not true and LECs cannot capture such states.

In this work, we show that the LECs can also determine the elementary excitations of the FQH states. For FQH phases where the CF construction or the Jack polynomial/CFT approach is applicable, the model wave functions obtained from LECs agree qualitatively, and semi-quantitatively, with the wave functions obtained using these traditional methods. The LEC approach can also be applied to novel FQH phases that do not lend themselves to a description in terms of CFs or a CFT. Furthermore, the LECs can describe excitations for certain phases, where the ground state is captured by a CFT, but there is no known description of the quasielectrons and neutral excitations. We shall present some examples of such phases below.

The construction of the model wave functions for elementary excitations using the LECs offers a new perspective on the nature of these excitations. In particular, we find that a set of LECs that defines the quasielectrons of one FQH phase could also define the ground state and the quasiholes of a different FQH phase. This not only sheds light on the relationship between different FQH phases but also reveals interesting links between phases that were previously believed to be unrelated. More specifically, we show strong evidence that the Gaffnian state~\cite{Simon07b} at $\nu=2/5$ is built from a particular type of quasielectrons of the Laughlin state at $1/3$. Similarly, the Laughlin state at $1/3$ can be viewed as arising from the condensation of the quasiholes of the Gaffnian state. Another example is the Fibonacci state in the Read-Rezayi series~\cite{Read99} which, following the LECs, can now be understood as consisting of a particular type of quasielectrons of the Moore-Read (MR) state~\cite{Moore91}. 

\begin{table}[h!]
\centering
\begin{tabular}{ |c|c|} 
 \hline
 $\hat n$&a single LEC given by a triplet $\{n,n_e,n_h\}$ \\ 
 \hline
 $n$     & number of flux quanta \\ 
 \hline
 $n_{e}$ & number of electrons \\ 
 \hline
 $n_{h}$ & number of holes \\ 
 \hline
 $\hat c$& constraint on Hilbert space, \\ 
 &e.g. $\hat c= \hat n_1, \hat c= \hat n_1\lor\hat n_2$ \\
 \hline
 $\mathcal H_{N_o,N_e}$ & Hilbert space on sphere with \\
 &$N_o$ orbitals and $N_e$ electrons\\
 \hline
  $\mathcal N_{N_o,N_e}$ & number of highest weight states in $\mathcal H_{N_o,N_e}$ \\
 \hline
 $\bar{\mathcal H}^{\hat c}_{N_o,N_e}$ & subset of $\mathcal H_{N_o,N_e}$ satisfying the constraint $\hat c$\\
 \hline
$\bar{\mathcal N}^{\hat c}_{N_o,N_e}$ & number of highest weight states in $\bar{\mathcal H}^{\hat c}_{N_o,N_e}$ \\
\hline
$\mathcal W^{\hat c}_{N_o,N_e}$ & subspace of $\bar{\mathcal H}^{\hat c}_{N_o,N_e}$ spanned by all highest\\
&weight states, with dimension $\bar{\mathcal N}^{\hat c}_{N_o,N_e}$\\
\hline
$\mathcal Q^{\hat t}_{N_o,N_e}$ & (dressed) quasielectron Hilbert space of type $\hat t$\\
\hline
\end{tabular}
\caption{Definition of various symbols used in the text.}
\label{t1}
\end{table}

We deploy the spherical geometry~\cite{Haldane83} for all our calculations. We note that the LECs, being physical constraints, can in principle, be applied to any geometry. However, the spherical geometry is the most convenient one since circular droplets at the north or south poles can be easily defined in this geometry. A nice feature of working on the sphere is that it is also the only geometry without a boundary, in which the topological shifts are well-defined even for finite systems. The two good quantum numbers on the sphere are the total orbital angular momentum $L$ and it's $z$-component $L_z$. For a rotationally invariant Hamiltonian, states with a given $L$ form a $(2L+1)$-degenerate multiplet with $-L\le L_z\le L$. The state with $L_z=L$ is defined to be the highest weight (HW) state in this multiplet. We only consider fully spin polarised quantum fluids here. Furthermore, we only look at the Hilbert space of a single Landau level (LL). We assume that the effects of LL mixing can be captured by more complicated dynamics within a single LL (e.g., by using three-body interactions).

The paper is organized as follows: In Sec.~\ref{qh} we show how the quasiholes of the FQH states can be constructed with the LEC formalism, and discuss the resulting bulk-edge correspondence between the ground state entanglement spectrum and the quasihole counting. Then, in Sec.~\ref{qene} we show how different types of quasielectrons and neutral excitations can be constructed with the LEC formalism. In Sec.~\ref{connection}, we show that the construction of quasielectron states using LECs reveals connections between the Gaffnian and the Laughlin state, as well as between the Fibonacci and the Moore-Read state. This leads to a new interpretation of the nature of these exotic topological states (Gaffnian and Fibonacci state) as consisting of charged excitations of simpler FQH systems (Laughlin and Moore-Read state). In Sec.~\ref{cftheory} we compare and contrast the elementary excitations constructed from the LEC formalism with those obtained from the composite fermion (CF) theory. We show that the excitations obtained from these two microscopically different approaches agree qualitatively in the region of filling factors between $\nu=1/3$ and $\nu=2/5$.   We conclude the paper in Sec.~\ref{summary} with a summary of our results and provide an outlook for the future.

\section{Quasihole state construction}\label{qh}
Let us first generalize the LEC construction of the FQH ground states~\cite{Yang19} to the quasihole states. The number of orbitals $N_{o}$ and the number of electrons $N_e$, of the ground states for incompressible quantum Hall systems satisfy the relation
\begin{equation}
N_{o}=\frac{q}{p}\left(N_e+S_e\right)-S_o. 
\label{expression}
\end{equation}
Here the filling factor $\nu=p/q$ ($p$ and $q$ are positive integers), and $S_e<p,S_o<q$ are integer topological shifts for the electrons and fluxes respectively. The total topological shift could be a rational number~\cite{Hutasoit16,Levin09,Mukherjee14}, requiring two integers, and here we use $S_e,S_o$ as the two integer indices explicitly. Extensive numerical evidence shows that the constraint imposed by one or a combination of LECs (denoted as $\hat c$) on a rotationally invariant quantum Hall fluid determines a set of topological indices $[p,q,S_e,S_o]$ satisfying the following commensurability conditions~\cite{Yang19}:
\begin{eqnarray}
&&N^d_{o}=\frac{q}{p}\left(N_e+S_e\right)-S_o\label{shift} \\
&&\bar{\mathcal N}^{\hat c}_{N^d_{o},N_e}=1,\quad\bar{\mathcal N}^{\hat c}_{N_{o}<N^d_{o},N_e}=0\label{ingap}
\end{eqnarray}
The commensurability conditions hold for all values of $N_e$ subject to the condition that $N_e+S_e=kp,k\ge 2$. Here, $\bar{\mathcal N}^{\hat c}_{N_{o},N_e}$ is the number of rotationally invariant ($L=0$) states of $N_e$ electrons in $N_o$ orbitals that  satisfy the constraint specified by $\hat c$. Thus given an allowed $N_e$, the unique highest density state contains $N^d_{o}$ orbitals (thus the superscript), is rotationally invariant and satisfies the constraint. Diagonalizing $L^2$ in the truncated Hilbert space (obtained by removing basis states that do not satisfy $\hat c$ at the north pole or south pole from the full set of basis states) results in the model wave function for the ground state of the FQH system indexed by $[p,q,S_e,S_o]$. Two prominent examples arising from a single LEC are~\cite{Yang19}: (i) Laughlin state~\cite{Laughlin83} at $\nu=1/\left(2n-1\right)$ which arises from $\hat c=\{n,1,n\}$ and corresponds to $[p,q,S_e,S_o]=[1,2n-1,0,2n-2]$, and (ii) The Read-Rezayi series~\cite{Read99} at $\nu=\left(n-1\right)/\left(n+1\right)$ which arises from $\hat c=\{n,n-1,n\}$ and corresponds to $[p,q,S_e,S_o]=[n-1,n+1,0,2]$.

From extensive numerical calculations, we find that $\bar{\mathcal N}^{\hat c}_{N_{o},N_e}$ is also the number of highest weight states in the corresponding truncated Hilbert space. If $\hat c$ corresponds to a particular commensurability condition $[p,q,S_e,S_o]$, then $\bar{\mathcal N}^{\hat c}_{N^d_{o},N_e}=1$ implies there is only one highest weight state even if we scan all possible $L_z$ sectors. Moreover, this unique highest weight state always occurs in the $L_z=0$ sector and is thus rotationally invariant. The quasihole states can be naturally obtained by fixing $N_e$ and letting $N_o>N_o^d$, which corresponds to inserting flux quanta into the highest density ground state. The quasihole states are the highest weight states (with $L=L_z$) that satisfy the same set of LECs as the ground state. Generically, the number of quasihole states at different values of $L$ are different from each other. We define $\mathcal W^{\hat c}_{N_{o},N_e}$ to be the subspace of $\bar{\mathcal H}^{\hat c}_{N_{o},N_e}$ spanned by all the highest weight states and we denote its dimension by $\bar{\mathcal N}^{\hat c}_{N_o,N_e}$. For $N_o=N_o^d$, $\mathcal W^{\hat c}_{N_o,N_e}$ is one-dimensional (ground state subspace) while for $N_o>N_o^d$, $\mathcal W^{\hat c}_{N_o,N_e}$ is the quasihole subspace.

For FQH states with a CFT description or a model Hamiltonian, the quasihole counting obtained from the LECs scheme matches exactly with the CFT or model Hamiltonian approach. For all states constructed from LECs, there is also an interesting bulk-edge correspondence between the counting of the ground state entanglement spectrum and the edge modes derived from bulk quasihole counting. This holds empirically for all the system sizes we have checked. Such bulk-edge correspondence is well-known in the CFT construction of the FQH states: the counting of levels in the entanglement spectrum of the ground state agrees with the edge state counting of the quasiholes constructed from the same CFT model~\cite{Li08, Chandran11}. The LEC construction indicates that such a correspondence holds even for states with no known CFT description, suggesting that the bulk-edge correspondence may be an intrinsic property of the algebraic structure of the truncated Hilbert space.
\begin{figure}[htb]
\includegraphics[width=\linewidth]{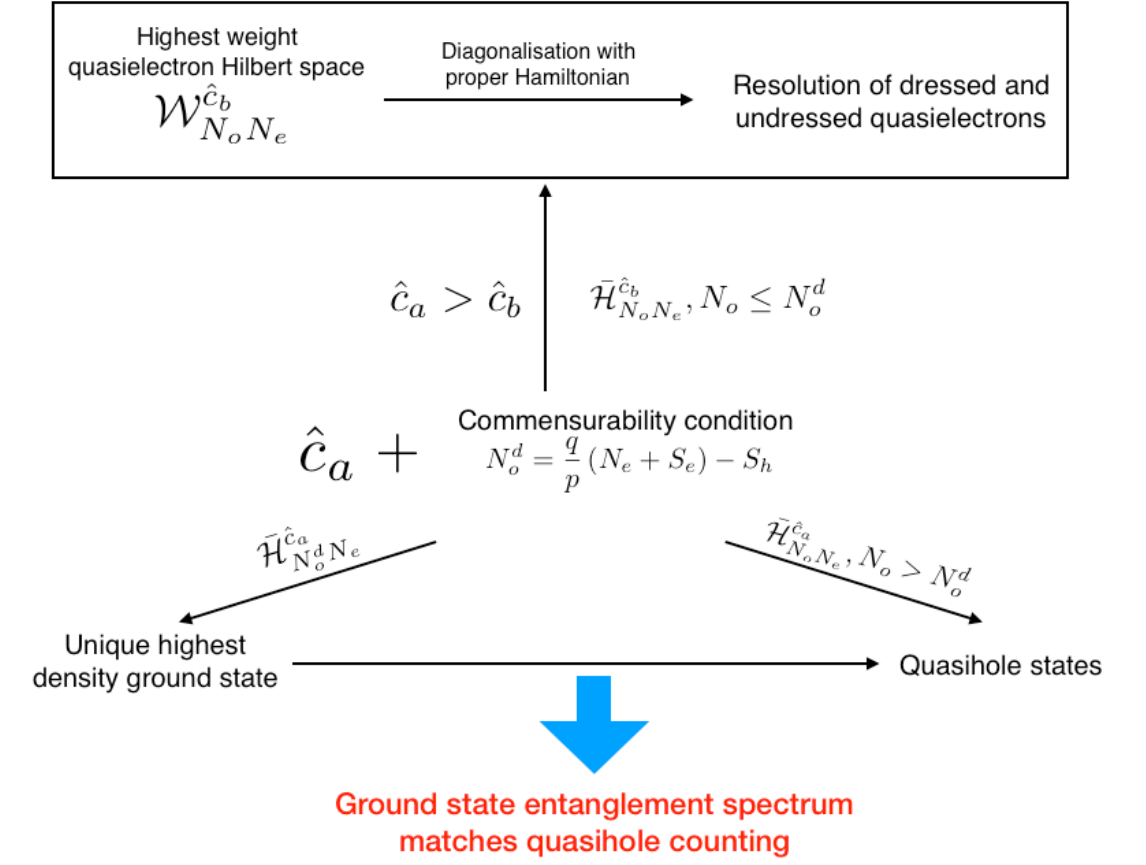}
\caption{Schematic relations between the LECs, the ground state, the quasihole manifold, and various types of quasielectron manifold. The relationship between the quasihole counting and the ground state entanglement spectrum is also illustrated. Here $\hat{c}_{a}>\hat{c}_{b}$ implies $\hat c_a$ is a more stringent constraint than $\hat c_b$.}
\label{fig1}
\end{figure}

\begin{figure}[H]
\includegraphics[width=\linewidth]{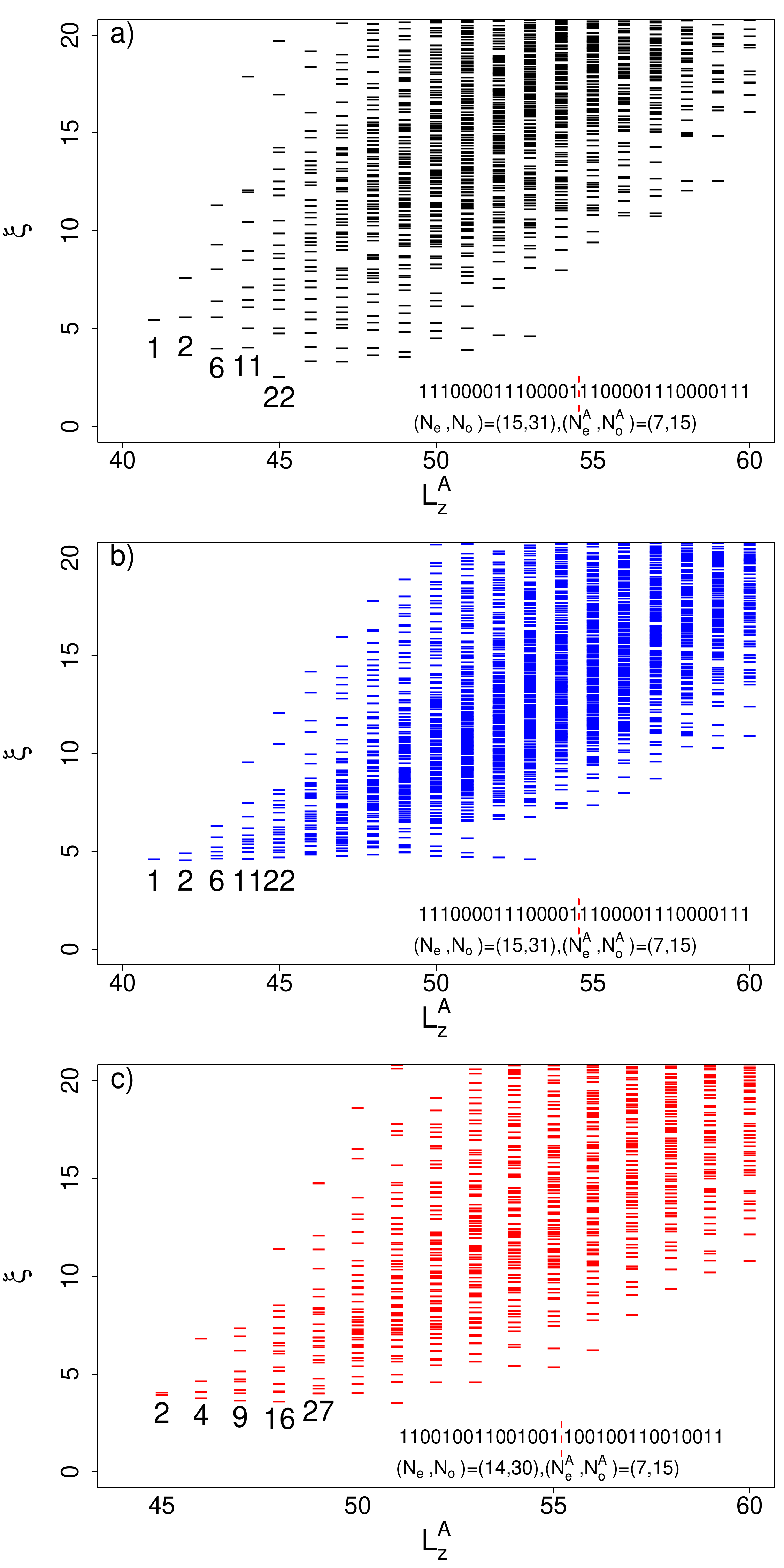}
\caption{a). The entanglement spectrum (ES) of the ground state corresponding to $[p,q,S_e,S_o]=[3,7,0,4]$ determined by $\hat c_1$ (see text for definition). b). The ES of the ground state corresponding to $[p,q,S_e,S_o]=[3,7,0,4]$ determined by $\hat c_2$ (see text for definition). c). The ES of the ground state corresponding to $[p,q,S_e,S_o]=[3,7,1,5]$ determined by $\hat c_3$ (see text for definition). The subsystem is given by $N_e^A$ and $N_o^A$ (partition of root configuration shown in the lower right corner). The numbers in the plots show the ES counting of the respective $L$ sector.}
\label{fig2}
\end{figure}

As an example, let us look at different topological phases at filling factor $\nu=3/7$. Consider two different constraints $\hat c_1=\{5,3,5\}$ and $\hat c_2=\{2,1,2\}\lor\{6,3,6\}$. The symbol $\lor$ in the latter means the quantum Hall fluids need to satisfy either $\{2,1,2\}$ \emph{or} $\{6,3,6\}$. Both $\hat c_1$ and $\hat c_2$ correspond to $[p,q,S_e,S_o]=[3,7,0,4]$. On one hand, the quasihole counting obtained from these two different constraints is identical. Moreover, the counting of levels in the entanglement spectra of the ground state wave functions obtained from these two constraints also agree with each other [see Fig.~\ref{fig2}a),\ref{fig2}b)]. On the other hand, the ground state wave functions obtained from $\hat c_1$ and $\hat c_2$ have vanishingly small overlap with each other. We conjecture that $\hat c_1$ and $\hat c_2$ realize different topological phases, which could be distinguished by analyzing their topological entanglement entropy (TEE)~\cite{Kitaev06a, Levin06} in the thermodynamic limit. At the moment, due to technical challenges, we do not have access to the ground state wave functions of these LECs for very large system sizes, which precludes a reliable extrapolation of their TEEs. A third topological phase at $\nu=3/7$ can be realised by $\hat c_3=\{3,2,3\}\land\{5,3,5\}$ [This state was first discovered by R. Thomale \emph{et al.} (unpublished).]. The symbol $\land$ implies that the quantum fluid needs to satisfy both $\{3,2,3\}$ \emph{and} $\{5,3,5\}$. The topological phase generated by $\hat c_3$ corresponds to $[p,q,S_e,S_o]=[3,7,1,5]$, and thus has different shifts, and different quasihole counting [leading to a different entanglement spectrum for the ground state (see Fig.~\ref{fig2}c))] compared to the phases obtained from $\hat c_1$ and $\hat c_2$. 

In all three cases, there are no apparent CFT descriptions of the FQH states, but the bulk-edge correspondence holds. In particular, the quasihole counting (and thus the edge state counting) is identical for $\hat c_1$ and $\hat c_2$, while $\hat c_3$ has a different counting. Empirically, this suggests that while different LECs determine different topological phases, the quasihole counting is only determined by the shifts and the filling factor.

\section{Quasielectron and neutral excitation construction}\label{qene}
We will now move on to the quasielectron and neutral excitations, and illustrate the construction for these states in the simplest case of the Laughlin state. Our methodology can be easily extended for other sets of LECs. It is instructive to first recall the construction of single quasielectron states from the Jack polynomial formalism. The starting point in the Jack formalism is the root configuration for the ground state, $100100100100\cdots$, where $\cdots$ denotes repeated patterns of $100$. A single quasielectron at the north pole can be added by flipping the leftmost $0$ to $1$ (adding one electron), and inserting two fluxes (or $0$'s). This results in a net addition of charge $(-e)/3$ ($-e$ is the charge of the electron) consistent with the charge of the quasielectron at $\nu=1/3$. Different ways of inserting two fluxes leads to different types of quasielectrons as listed below:
\begin{eqnarray}\label{qe}
&&\d{1}\d{1}0\textsubring{0}01001001001001\cdots\quad\left(5,2\right)\text{Type}\quad L=N_e/2\nonumber\\
&&\d{1}\d{1}0010\textsubring{0}01001001001\cdots\quad\left(6,3\right)\text{Type}\quad L=N_e/2+1\nonumber\\
&&\d{1}\d{1}0010010\textsubring{0}01001001\cdots\quad\left(7,4\right)\text{Type}\quad L=N_e/2+2\nonumber\\
&&\vdots
\end{eqnarray} 
We name the quasielectron types as the $\left(5,2\right)\text{Type}$ or $\left(6,3\right)\text{Type}$ etc. for reasons that would become apparent later. The solid and empty circles below the root configuration indicate the locations of $-e/3$ quasiparticles and $+e/3$ quasiholes respectively. Here, the quasiparticles (or quasiholes) are located in regions where three consecutive orbitals in the root configuration accommodate more (or less) than a single electron. While a quasihole is an elementary excitation with charge $+e/3$, a quasielectron (not a quasiparticle) is the elementary excitation with charge $-e/3$. The quasielectron has a nontrivial internal structure as a bound state of two quasiparticles and one quasihole. Each of the quasielectron states only consists of the squeezed basis~\cite{Bernevig09b} from the respective root configurations shown in Eq.~(\ref{qe}). They can be uniquely determined by the highest weight condition, with the constraint that the state relaxes back to the Laughlin ground state away from the north pole. Such quasielectron states are identical to the composite fermion construction, where different types of quasielectrons correspond to adding a single composite fermion in different CF levels~\cite{Majumder09}.

The basis squeezed from the $\left(5,2\right)\text{Type}$ root configuration manifestly satisfies $\hat c_\alpha=\{2,1,2\}\lor\{5,2,5\}$, which is the set of LECs that define the Gaffnian ground state and its quasiholes~\cite{Yang19}. Thus the $\left(5,2\right)\text{Type}$ quasielectron state can also be understood as a condensation of the Gaffnian quasiholes (see Sec.~\ref{connection}). The $\left(5,2\right)\text{Type}$ quasielectron is a charged excitation of the Laughlin state (defined by $\{2,1,2\}$~\cite{Yang19}) since the quasielectron at the north pole breaks the $\{2,1,2\}$ constraint. However, the $\left(5,2\right)\text{Type}$ quasielectron still satisfies $\hat c_\alpha$. This justifies the use of our terminology for the various types of quasielectrons in Eq.~(\ref{qe}). In fact, if we impose $\hat c_\alpha$ on the entire Hilbert space of $L_z=N_e/2$ states, together with the highest weight condition, we obtain a large number of Gaffnian quasihole states. We can now interpret these states as containing a single quasielectron and potentially multiple neutral excitations (a neutral excitation is composed of a quasielectron and a quasihole) on top of the Laughlin state at $\nu=1/3$. These different states can be resolved by diagonalizing the highest weight subspace with the $V_1$ Haldane pseudopotential interaction (referred to as $V_{1}$ Hamiltonian from here on)~\cite{Haldane83}.

The two-quasielectron (of $\left(5,2\right)\text{Type}$) states can be constructed by imposing $\hat c_\alpha$ on the Hilbert space of $N_o=3N_e-4$, and obtaining the highest weight subspace in different $L^2$ sectors. Analyzing the ground state energy by diagonalizing the $V_1$ Hamiltonian within each subspace, we can extract wave functions and the counting of the two-quasielectron states. This procedure generalises to multi-quasielectron states, where for $n$ quasielectrons, the target Hilbert space has $N_o=3N_e-2-n$. Moreover, in the Hilbert space of $N_o=3N_e-2$, where no fluxes or electrons are added, we can use the same $\hat c_\alpha$ to extract the neutral excitations. In particular, the magnetoroton mode~\cite{Girvin85, Girvin86} is formed by the highest weight state in the sectors with $L_{z}=2,3,\cdots, N_{e}$.
\begin{figure*}[h]
\centering
\includegraphics[width=15 cm]{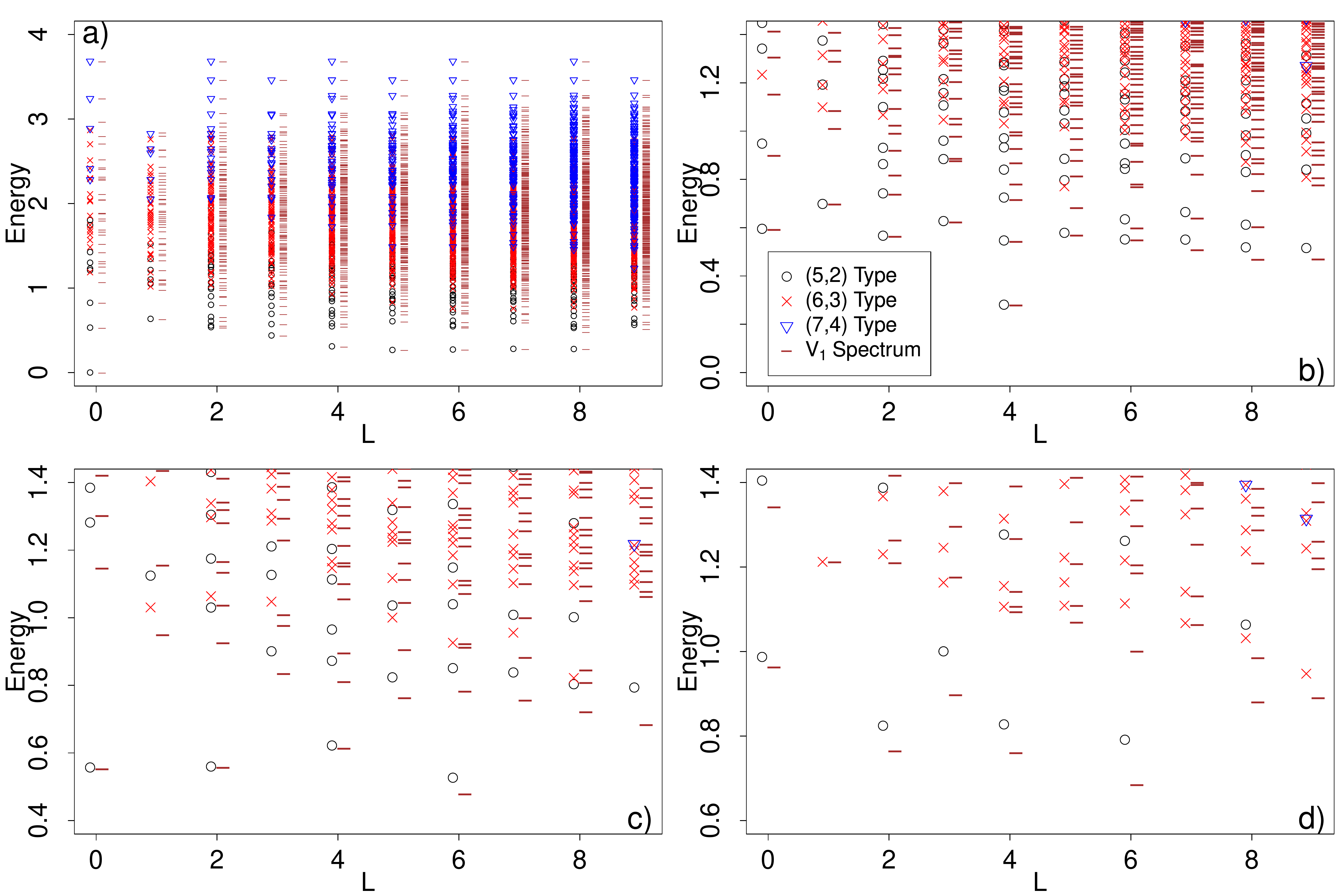}
\caption{Energy spectra resolved by the model Hamiltonian containing only $V_1$ pseudopotential interaction, within each excitation type (e.g. $\mathcal H^{\text{qe}}_{\left(5,2\right)},\mathcal H^{\text{qe}}_{\left(6,3\right)}$, etc.). The energy spectra of the full Hilbert space with the same Hamiltonian (allowing coupling between different excitation types, e.g. between $\mathcal H^{\text{qe}}_{\left(5,2\right)}$ and $\mathcal H^{\text{qe}}_{\left(6,3\right)}$, etc.) are listed immediately on the right for comparison. a). $N_e=8, N_o=22$, so only neutral excitations are present; b). $N_e=8, N_o=21$, all states containing a single quasielectron (may be dressed by neutral excitations); c). $N_e=8, N_o=20$, all states containing two quasielectrons (may be dressed); d). $N_e=8, N_o=19$, all states containing three quasielectrons (may be dressed). For b), c), d), only the low-lying part of the spectrum is shown. The legend in b) is shared by all panels.}
\label{fig3}
\end{figure*} 

More generally, with $\hat c_\alpha=\{2,1,2\}\lor\{5,2,5\}$, the Hilbert space $\mathcal W^{\hat c_\alpha}_{N_o,N_e}$ can be interpreted either as the Gaffnian quasihole manifold (when $N_o>5N_e/2-3$), or as the Laughlin $\left(5,2\right)\text{Type}$ quasielectron manifold (when $N_o<3N_e-2$). Here, the Laughlin $\left(5,2\right)\text{Type}$ quasielectron manifold is defined as the one spanned by states that contain either undressed $\left(5,2\right)\text{Type}$ quasielectrons, or dressed $\left(5,2\right)\text{Type}$ quasielectrons (e.g. dressed by neutral excitations if $N_o=3N_e-2$, or dressed by quasielectrons and neutral excitations if $N_o<3N_e-2$).

Each species of quasielectrons in Eq.~(\ref{qe}) is similarly defined by its respective LECs. The $\left(6,3\right)\text{Type}$ is defined by $\hat c_\beta=\{2,1,2\}\lor\{6,3,6\}$, while the $\left(7,4\right)\text{Type}$ is defined by $\hat c_\gamma=\{2,1,2\}\lor\{7,4,7\}$, and so on. We note that $\mathcal W^{\hat c_\alpha}_{N_o,N_e}\subseteq \mathcal W^{\hat c_\beta}_{N_o,N_e}\subseteq \mathcal W^{\hat c_\gamma}_{N_o,N_e}$, since the constraint $\hat c_\alpha$ is stricter than $\hat c_\beta$, which in turn is stricter than $\hat c_\gamma$. All such Hilbert spaces can be explicitly computed numerically for allowed values of $N_e$ and $N_o$. Let us define $\mathcal Q^{\left(5,2\right)}_{N_o,N_e}\coloneqq\mathcal W^{\hat c_\alpha}_{N_o,N_e}$ as the Hilbert space spanned by all states containing only $\left(5,2\right)\text{Type}$ quasielectrons. Similarly, $\mathcal Q^{\left(6,3\right)}_{N_o,N_e}$ is spanned by states that contain at least one $\left(6,3\right)\text{Type}$ quasielectron. Note that $\mathcal Q^{\left(6,3\right)}_{N_o,N_e}$ can contain some $\left(5,2\right)\text{Type}$ quasielectrons, and these $\left(5,2\right)\text{Type}$ quasielectrons may or may not be dressed with $\left(5,2\right)\text{Type}$ or $\left(6,3\right)\text{Type}$ neutral excitations. Numerically, $\mathcal Q^{\left(6,3\right)}_{N_o,N_e}$ can be easily determined by projecting out $\mathcal W^{\hat c_\alpha}_{N_o,N_e}$ from $\mathcal W^{\hat c_\beta}_{N_o,N_e}$. We can also define $\mathcal Q^{\left(7,4\right)}_{N_o,N_e}$ analogously by projecting out $ \mathcal W^{\hat c_\beta}_{N_o,N_e}$ from $ \mathcal W^{\hat c_\gamma}_{N_o,N_e}$ and so on. Therefore we have the following relations:
\begin{eqnarray}\label{relation}
&&\mathcal Q^{\left(6,3\right)}_{N_o,N_e}\subseteq \mathcal W^{\hat c_\beta}_{N_o,N_e}, \mathcal Q^{\left(6,3\right)}_{N_o,N_e}\perp \mathcal W^{\hat c_\alpha}_{N_o,N_e}.\nonumber\\
&&\mathcal Q^{\left(7,4\right)}_{N_o,N_e}\subseteq \mathcal W^{\hat c_\gamma}_{N_o,N_e}, \mathcal Q^{\left(7,4\right)}_{N_o,N_e}\perp \mathcal W^{\hat c_\beta}_{N_o,N_e}.\nonumber\\
&&\qquad\qquad\qquad\qquad\vdots,
\end{eqnarray}
where $\perp$ denotes the two spaces are orthogonal. In this way, the entire Hilbert space of the Laughlin phase can be systematically organized. Each quasielectron manifold $\mathcal Q^{\hat t}_{N_o,N_e}$ (where $\hat t$ denotes the LEC type) are physically distinct and orthogonal to each other. This is because different types of quasielectrons have different intrinsic angular momentum, and thus, in principle, could be experimentally distinguished. 

In Fig.~\ref{fig3} we show how the entire Hilbert space could be organized into states containing different types of excitations. While such an organization can be carried out for any Hilbert space (determined by $N_e$ and $N_o$), independent of the microscopic Hamiltonian, it is most relevant physically if the Hamiltonian is dominated by $V_1$ (so that we are in the Laughlin phase). It is important to note that the excitation Hilbert spaces are determined algebraically without resorting to any local operators. Here we use Hamiltonians to show such algebraically determined Hilbert spaces are physically relevant for realistic interactions. Taking $\mathcal Q^{\left(5,2\right)}_{N_o,N_e}$ as an example, all states in this Hilbert space contain $\left(5,2\right)\text{Type}$ quasielectron(s) and possibly some Laughlin quasiholes. The microscopic Hamiltonian crucially decides whether these quasielectrons can be effectively treated as ``elementary particles". The $V_1$ Hamiltonian can be diagonalised within $\mathcal Q^{\left(5,2\right)}_{N_o,N_e}$ to resolve states containing different number of quasielectrons. In a multi-quasielectron state, interactions between quasielectrons, as well as between quasielectrons and quasiholes, can make it difficult to ascertain the precise number of quasielectrons in a particular state. This is because the variational energies are no longer just an integer multiple of the single quasielectron creation energy. The deviation, however, is small if the quasielectrons are far away from each other.

The magnetoroton mode (one $\left(5,2\right)\text{Type}$ quasielectron dressed by one quasihole) can be clearly seen in Fig.~\ref{fig3}a) for a neutral system. For large momenta when the quasielectron is well-separated from the quasihole, the creation energy of a single quasielectron is roughly equal to the variational energy of the state (since a quasihole far away from other excitations costs negligible energy with the $V_1$ Hamiltonian). This variational energy also matches a single undressed quasielectron state in Fig.~\ref{fig3}b), as the lowest energy state at $L=N_e/2$. All other higher energy states contain one single quasielectron dressed by $\left(5,2\right)\text{Type}$ neutral excitations and therefore contain multiple quasielectrons. In Fig.~\ref{fig3}c) and Fig.~\ref{fig3}d), the lowest energy states contain two and three undressed quasielectrons respectively.

A similar analysis can be done by diagonalizing the $V_1$ Hamiltonian within $\mathcal Q^{\left(6,3\right)}_{N_o, N_e}$ or sectors of other types of quasielectrons, and states containing undressed quasielectrons of different types can also be determined. With $V_1$ Hamiltonian, the creation energies of these quasielectrons are higher than that of the $\left(5,2\right)\text{Type}$. Additional branches of neutral excitations discovered in Ref.~\cite{Yang19} can also be constructed just like the magnetoroton mode in $\mathcal Q^{\left(5,2\right)}_{N_o,N_e}$. The key message here is that quasielectrons can be treated as weakly interacting particles for short-range Hamiltonians dominated by $V_1$. Also, different types of undressed quasielectrons have different intrinsic angular momenta since the highest weight states live in different $L^2$ sectors~\cite{Yang14}. Thus a rotationally invariant Hamiltonian cannot couple a single quasielectron of one type to another quasielectron of a different type.

In Fig.~\ref{fig3} we also show the exact energy spectra of different systems (specified by $N_e$ and $N_o$, with $V_1$ Hamiltonian), comparing them with the energy levels (we denote as variational spectra) obtained from diagonalising the same $V_1$ Hamiltonian in each quasielectron manifold $\mathcal Q^{\hat t}_{N_o,N_e}$ separately. In each $L$ sector, the low-lying states predominantly coming from $\mathcal Q^{\left(5,2\right)}_{N_o, N_e}$ match very well with the exact spectra, both in terms of the variational energies and wave function overlaps ($>99\%$). The same is true for the highest energy states, which consist of different types of quasielectrons depending on the system size and the orbital angular momentum. The mismatch between the exact spectra and variational spectra is because the Hamiltonian couples different $\mathcal Q^{\hat t}_{N_o, N_e}$, at least for finite systems with $V_1$ interaction. Even though a single undressed quasielectron of one type cannot scatter into another type of undressed quasielectron because they have different quantum numbers, it can scatter into other types of \textit{dressed} quasielectrons. Matrix elements coupling multiple quasielectrons of different types are non-zero if they carry the same quantum number. Thus with realistic interactions, only a dilute gas of quasielectrons can be resolved in terms of the different quasielectron types. If the interaction is dominated by $V_{1}$ the low lying quasielectrons are of the $\left(5,2\right)\text{Type}$ (if they are present). On the other hand, in large $L$ sectors, quasielectrons of other types (e.g. the $\left(6,3\right)\text{Type}$) are the low lying ones (since the $\left(5,2\right)\text{Type}$ is absent). These other types of quasielectrons can be experimentally detected as excitations with large angular momentum transfer from the probing particles (e.g. neutrons or photons) to the FQH fluid~\cite{Pinczuk93, Kukushkin09}.

An important comment is in order here about the need for interaction input to resolve microscopic wavefunctions of the quasielectron and neutral excitation states. The LEC formalism determines the manifolds or the sub-Hilbert spaces (spanned by highest weight states in the truncated Hilbert space) of different types of quasielectrons that differ in their universal topological properties. Such sub-Hilbert spaces are universal, with their properties determined without any interaction input. The microscopic wavefunctions and their variational energies, on the other hand, are non-universal and depend on the details of the microscopic Hamiltonian. Understanding the universal topological properties of these excitations with the LEC formalism could allow a systematic analysis for the robustness of the topological properties with realistic interactions. To this end, in Sec.~\ref{cftheory}, we have compared the states obtained from the LEC formalism with the exact states of the LLL Coulomb interaction.

\section{Connections between different FQH phases}
\label{connection}
We have shown that a set of LECs can define both the ground state and quasihole states of one FQH phase, and at the same time define the quasielectron and neutral excitations manifolds of a different FQH phase. This leads to a new perspective on the nature of many topological phases, as well as the connections between them. Understanding the ground state of one FQH phase as the quantum fluid of the excitations of a topologically different FQH phase is a common perspective in the hierarchy~\cite{Haldane83, Halperin84} and Jack polynomial approach~\cite{Bernevig09b}. The interactions between quasielectrons and quasiholes result from the Coulomb interaction between electrons. In many cases~\cite{Lee01, Lee02, Balram16c, Balram17b}, such interactions are much weaker than the Coulomb interactions. Thus the excitations themselves serve as more convenient ``elementary particles" in describing other FQH phases, as compared to using electrons. Here in the LEC formalism, the quasielectron/quasihole manifolds are universal and completely determined by the Hilbert space algebra. The identification of these sub-Hilbert space between different FQH phases not only reproduces well-known connections between FQH phases but also reveals unexpected connections between exotic FQH phases.

As an example, $\hat c_\alpha=\{2,1,2\}\lor\{5,2,5\}$ defines the Gaffnian state and its quasiholes~\cite{Yang19}, as well as the $\left(5,2\right)\text{Type}$ quasielectron and neutral excitations (the constituent quasielectron of the neutral excitation is of the $\left(5,2\right)\text{Type}$) of the Laughlin state at $\nu=1/3$. Thus this family of quasielectrons and neutral excitations can also be understood as the Gaffnian quasihole states when the quasiholes are uniformly spaced far away from the local defect at which $\hat n=\{2,1,2\}$ is violated. Similarly, the Laughlin state can also be understood as a condensation of the Gaffnian quasiholes. This perspective is familiar in the Jack polynomial formalism (with the root configurations), and in the hierarchical approach where the Laughlin state (and its excitations) are the condensate of the $2/5$ quasiholes~\cite{Haldane83, Halperin84}. The relationship between the Gaffnian state, the Laughlin state, and the Abelian Jain $2/5$ state has also been discussed in the literature~\cite{Yang19b,yang2020}.

The Gaffnian ground state is also closely linked to the $\left(5,2\right)\text{Type}$ quasielectrons of the Laughlin states, which is very much reminiscent of the hierarchical or the CF construction. We shall report in detail elsewhere on the connections between the Gaffnian and CF states at $\nu=2/5$. Similarly, the $\left(6,3\right)\text{Type}$ quasielectrons at $\nu=1/3$ establishes a connection between the FQH phase at $\nu=1/3$ and $\nu=3/7$. This is because the $\left(6,3\right)\text{Type}$ quasielectrons and the $\nu=3/7$ ground state (together with its quasiholes) are all defined by $\hat c_\beta=\{2,1,2\}\lor\{6,3,6\}$ (which is equivalent to $\{2,1,2\}\lor\{5,2,5\}\lor\{6,3,6\}$). We find that the $\nu=3/7$ ground state determined by the $\hat c_\beta$ LEC has a high overlap with the Jain $3/7$ state as well as the LLL Coulomb ground state [see Table~\ref{tab:overlaps_n_0_LL_3_7_Jain_LEC}].  The Fock space representations of the Jain $3/7$ state [see Eq.~(\ref{eq_Jain_CF_wf})] for the systems of $N_{e}=9$ and $12$ electrons were obtained by a brute-force direct projection to the LLL~\cite{Dev92}. We have not been able to obtain the next Jain $3/7$ state of $N_{e}=15$ electrons in the Fock space since the computation is very time and resource consuming. On the other hand, we have been able to evaluate the $3/7$ LEC states for up to $N_{e}=18$ electrons since the LEC states can just be obtained by diagonalizing the $L^{2}$ operator in the constraint-satisfying Hilbert space.
\begin{table}[h]
\centering
\begin{tabular}{|c|c|c|c|c|}
\hline
$N_{e}$ & $N_{o}$ & $|\langle \Psi^{0{\rm LL}}_{3/7}|\Psi^{\rm LEC}_{3/7} \rangle|$ &  $|\langle \Psi^{\rm Jain}_{3/7}|\Psi^{\rm LEC}_{3/7} \rangle|$ & $|\langle \Psi^{0{\rm LL}}_{3/7}|\Psi^{\rm Jain}_{3/7} \rangle|$  \\ \hline
9   & 17   		&0.9851    & 0.9866 &	0.9994~\cite{Wu93,Jain07}	\\ \hline
12  & 24  		&0.9652    & 0.9720 &	0.9988~~~~~~~~~	\\ \hline
15  & 31 	 	&0.9509    & -      &	  -   	\\ \hline
18  & 38 	 	&0.9430    & -      &	  -   	\\ \hline
\end{tabular}
\caption{\label{tab:overlaps_n_0_LL_3_7_Jain_LEC} Overlaps of the ground state for $N_{e}$ electrons in $N_{o}$ orbitals of the spherical geometry at the Jain $3/7$ flux in the $n=0$ Landau level (obtained by exact diagonalization), $\Psi^{0{\rm LL}}_{3/7}$, with the LEC state, $\Psi^{\rm LEC}_{3/7}$, generated by the constraint $\hat c_\beta=\{2,1,2\}\lor\{6,3,6\}$ and Jain $3/7$ state, $\Psi^{\rm Jain}_{3/7}$.}
\end{table}

\begin{figure*}[h]
\includegraphics[width=15 cm]{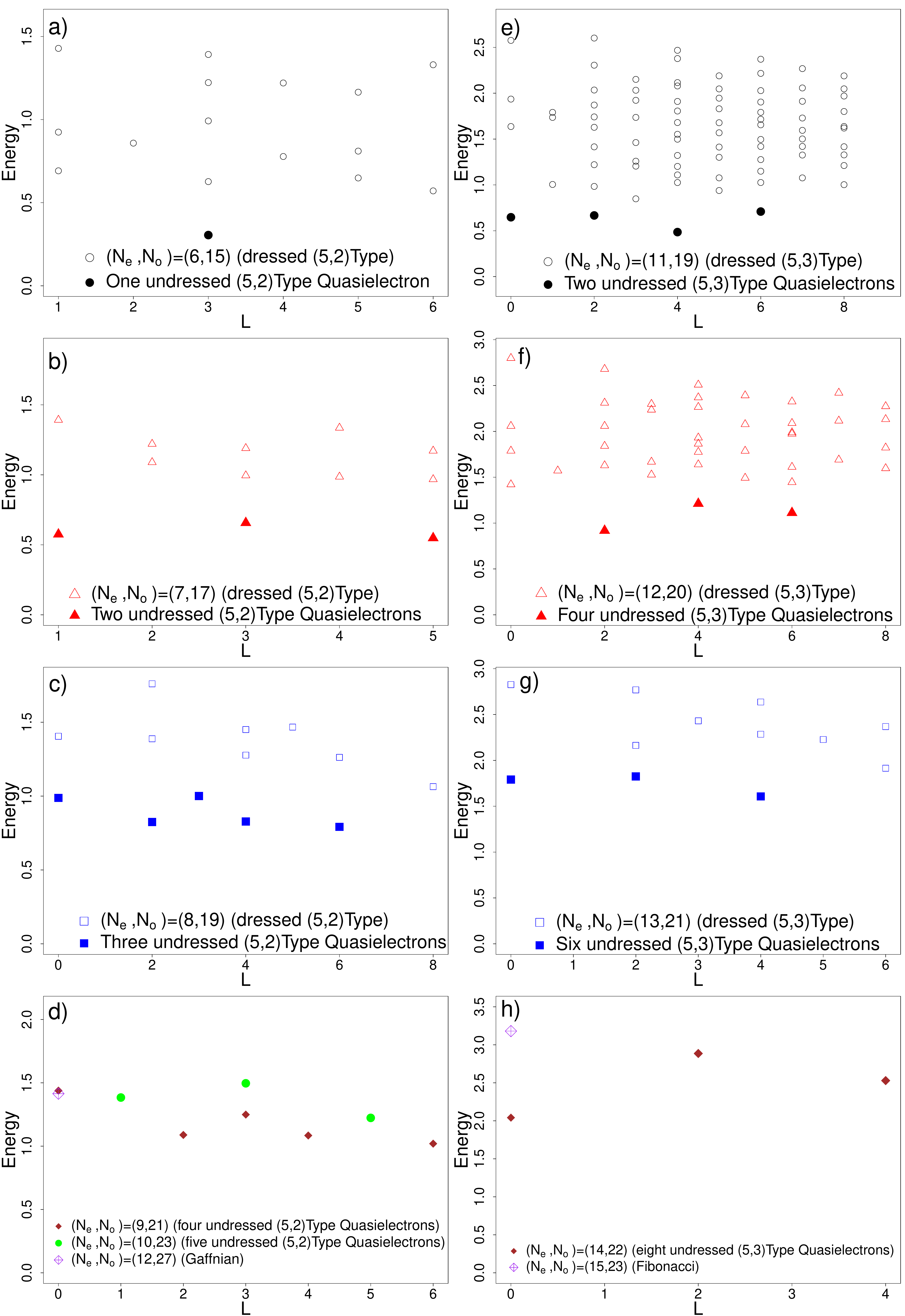}
\caption{a)$\sim$ d): Laughlin $\left(5,2\right)\text{Type}$ quasielectron states for different system sizes. There are no dressed quasielectrons for system sizes in d)., and there are no $\left(5,2\right)\text{Type}$ quasielectrons for $11$ electrons and $25$ orbitals. e)$\sim$ h). Moore-Read $\left(5,3\right)\text{Type}$ quasielectron states for different system sizes.}
\label{fig4}
\end{figure*}

The fact that the Gaffnian state can be reinterpreted as the Laughlin $\left(5,2\right)\text{Type}$ quasielectron state is illustrated in Fig.~\ref{fig4}a). Starting with the Laughlin ground state at $\nu=1/3$ with $N_e$ electrons and $N^d_o=3N_e-2$ orbitals, we can add $n_{\text{qe}}$ number of $\left(5,2\right)\text{Type}$ quasielectrons by adding $n_{\text{qe}}$ electrons and $2n_{\text{qe}}$ orbitals. Diagonalising the respective $\mathcal Q^{\left(5,2\right)}_{N_o,N_e}$ Hilbert space with the $V_1$ Hamiltonian, states containing $n_{\text{qe}}$ undressed quasielectrons can be resolved. In particular, when $n_{\text{qe}}=N_e+2$, there is only one highest weight state in $\mathcal Q^{\left(5,2\right)}_{N_o,N_e}$, which is translationally invariant. It turns out that this state is precisely the model Gaffnian state. 

To see how such connections go beyond the Laughlin state, we look at the Moore-Read (MR) state~\cite{Moore91} at $\nu=2/4$ with $[p,q,S_e,S_o]=[2,4,0,2]$. In the LEC construction, the MR state as well as its (non-Abelian) quasiholes can be uniquely defined by $\hat c_m=\{3,2,3\}$. Like the Laughlin state, there can be different types of quasielectrons (and thus neutral excitations) for the MR state. These arise from different ways of inserting the flux, after converting a hole into an electron at the north pole. The simplest type of quasielectron is of the $\left(5,3\right)\text{Type}$. Its root configuration and that of the corresponding neutral excitations are as follows:
\begin{eqnarray}
&&\left(5,3\right)\text{Type quasielectron}\nonumber\\
&&\qquad\qquad\qquad\d{1}\d{1}100110011001100\cdots\quad\nonumber\\
&&\left(5,3\right)\text{Type neutral excitations (magnetoroton mode)}\nonumber\\
&&\qquad\qquad\qquad\d{1}\d{1}1\textsubring{0}01\textsubring{0}011001100\cdots\nonumber\\
&&\qquad\qquad\qquad\d{1}\d{1}1\textsubring{0}010101\textsubring{0}01100\cdots\nonumber\\
&&\qquad\qquad\qquad\d{1}\d{1}1\textsubring{0}01010100101\textsubring{0}\cdots\nonumber\\
&&\qquad\qquad\qquad\qquad\qquad\vdots\nonumber\\
&&\left(5,3\right)\text{Type neutral excitations (neutral Fermion mode)}\nonumber\\
&&\qquad\qquad\qquad\d{1}\d{1}1\textsubring{0}\textsubring{0}0110011001100\cdots\nonumber\\
&&\qquad\qquad\qquad\d{1}\d{1}1\textsubring{0}0101\textsubring{0}011001100\cdots\nonumber\\
&&\qquad\qquad\qquad\d{1}\d{1}1\textsubring{0}01010101\textsubring{0}01100\cdots\nonumber\\
&&\qquad\qquad\qquad\qquad\qquad\vdots
\end{eqnarray}

For even and odd number of electrons, the neutral excitations form the well-known magnetoroton mode and the neutral fermion mode respectively. These quasielectron and neutral excitations can be completely defined by $\hat c_{m}=\{3,2,3\}\lor\{5,3,5\}$. It turns out $\hat c_{m}$ also identifies the Fibonacci state at $\nu=3/5$, which is the next state after the MR state in the Read-Rezayi sequence~\cite{Read99}. Starting with the MR ground state at $\nu=1/2$ with $N_e$ electrons and $N^d_o=2N_e-2$ orbitals, every time we add one electron and one flux (orbital), we are adding two quasielectrons, each with charge $-e/4$. We continue this process by adding two quasielectrons at a time. The undressed  $\left(5,3\right)\text{Type}$ quasielectrons can be resolved by first obtaining the highest weight quasielectron Hilbert space $\mathcal Q^{\left(5,3\right)}_{N_o,N_e}$ with $\hat c_{m}$, then diagonalising within this Hilbert space with the model three-body Hamiltonian for the MR state. After adding $N_e$ electrons and $N_e$ orbitals, the only highest weight state in $\mathcal Q^{\left(5,3\right)}_{N_o,N_e}$ is the translationally invariant Fibonacci state, or the state containing $2N_e$ quasielectrons of the $\left(5,3\right)\text{Type}$ [see Fig.~\ref{fig4}b)].

\section{Comparison of the composite fermion and LEC constructions}\label{cftheory}
In this section, we compare the wave functions obtained from the composite fermion (CF) and LEC constructions in detail. We show that for quasiholes and undressed quasielectrons, these two theories, which are microscopically different, lead to model wave functions that agree with each other qualitatively and semi-quantitatively. For the sake of completeness, we will first provide a primer on the CF theory. 

A vast majority of the LLL FQHE phenomena is captured in terms of emergent topological particles called composite fermions, which are bound states of electrons and an even number $(2p)$ of quantized vortices~\cite{Jain89}. CF theory postulates that a system of interacting electrons at filling factor $\nu=\nu^{*}/(2p\nu^{*}\pm 1)$ can be mapped onto a system of weakly interacting composite fermions at a filling factor $\nu^{*}$. In particular, integer filling of CF-LLs (termed $\Lambda$Ls), i.e $\nu^{*}=n$, leads to FQHE of electrons at $\nu=n/(2pn\pm 1)$. The mapping to integer quantum Hall effect (IQHE) leads to the following Jain wave functions for interacting electrons in the LLL~\cite{Jain89}:
\begin{equation}
\Psi^{\alpha}_{\nu=\frac{\nu^{*}}{2p\nu^{*}\pm 1}} =
\mathcal{P}_{\rm LLL} \Phi_{1}^{2p}\Phi^{\alpha}_{\pm\nu^{*}}. 
\label{eq_Jain_CF_wf}
\end{equation}
Here $\alpha$ labels the different eigenstates, $\Phi_{\nu^{*}}$ is the Slater determinant wave function of electrons at $\nu^{*}$ (with $\Phi_{-\nu^{*}}=\overline{[\Phi_{\nu^{*}}]}$, where overline denotes complex conjugation) and $\mathcal{P}_{\rm LLL}$ implements projection to the LLL. Throughout this work we carry out projection to the LLL using the Jain-Kamilla method~\cite{Jain97,Jain97b}, details of which can be found in the literature~\cite{Moller05,Jain07,Davenport12,Balram15a}.

The quasihole and quasielectron excitations of the FQH systems are obtained as composite fermion hole (CFH) and composite fermion particle (CFP) respectively. A CFH is a missing composite fermion in an otherwise full $\Lambda$L, while a CFP is a composite fermion in an otherwise empty $\Lambda$L. Wave functions of the CFP and CFH can be constructed along the lines of Eq.~(\ref{eq_Jain_CF_wf}) using the analogy to the particle and hole excitations in the corresponding IQH state. The CF theory is in excellent qualitative as well as quantitative agreement with exact diagonalization studies of the LLL Coulomb problem for both the ground states as well as the excitations~\cite{Dev92, Jain97, Jain07, Balram13, Balram20, Balram20a}. 

We shall focus our attention on the simplest FQH state at $\nu=1/3$, which in the CF theory maps to a $\nu^{*}=1$ state. The ground state wave function of Eq.~(\ref{eq_Jain_CF_wf}) for this case reduces to the Laughlin state~\cite{Laughlin83}. Furthermore, the wave function of the CFH at $\nu=1/3$ is identical to that of the Laughlin quasihole~\cite{Laughlin83} which is identical to the quasihole obtained from the LEC construction. Thus, to compare the CF and LEC constructions we shall mainly focus on the CFP. The wave function of the CFP at $\nu=1/3$ is not identical to that of the Laughlin quasielectron~\cite{Laughlin83}. However, the CFP and the Laughlin quasielectron have high overlaps with each other for small systems~\cite{Jain07} and are believed to describe the same excitation. The orbital angular momentum $L$ of a single CFP at $\nu=1/3$ is $L^{\rm CFP}_{1/3}=N/2$~\cite{Jain07,Balram16d}. States consisting of multiple CFPs at $\nu=1/3$ are constructed using the analogy to multiple particle states at $\nu^{*}=1$. The orbital angular momenta of states consisting of multiple CFPs can be ascertained by adding the angular momenta of the constituent CFPs.

\begin{figure*}[h]
\includegraphics[width=0.32\textwidth]{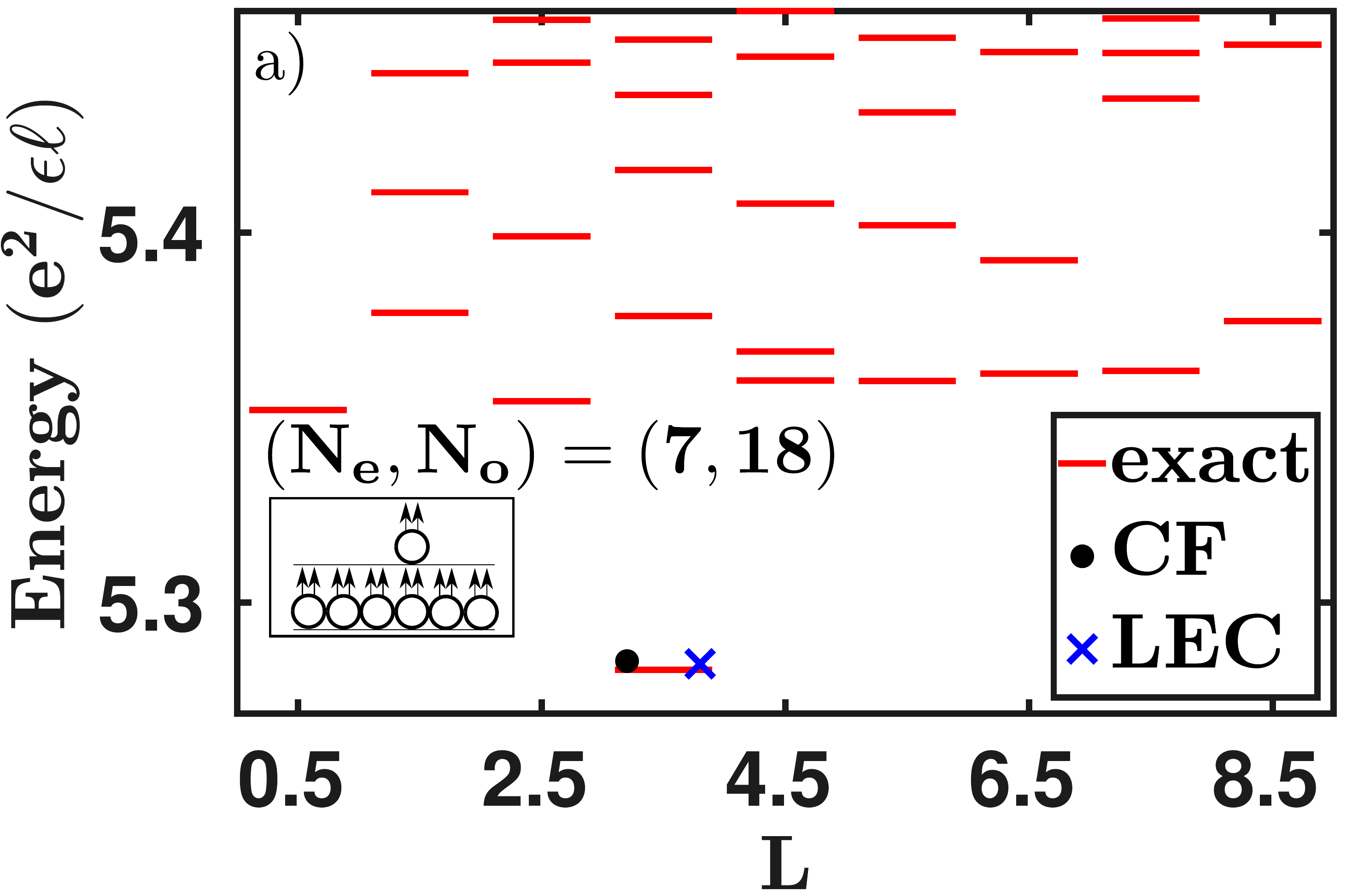}
\includegraphics[width=0.32\textwidth]{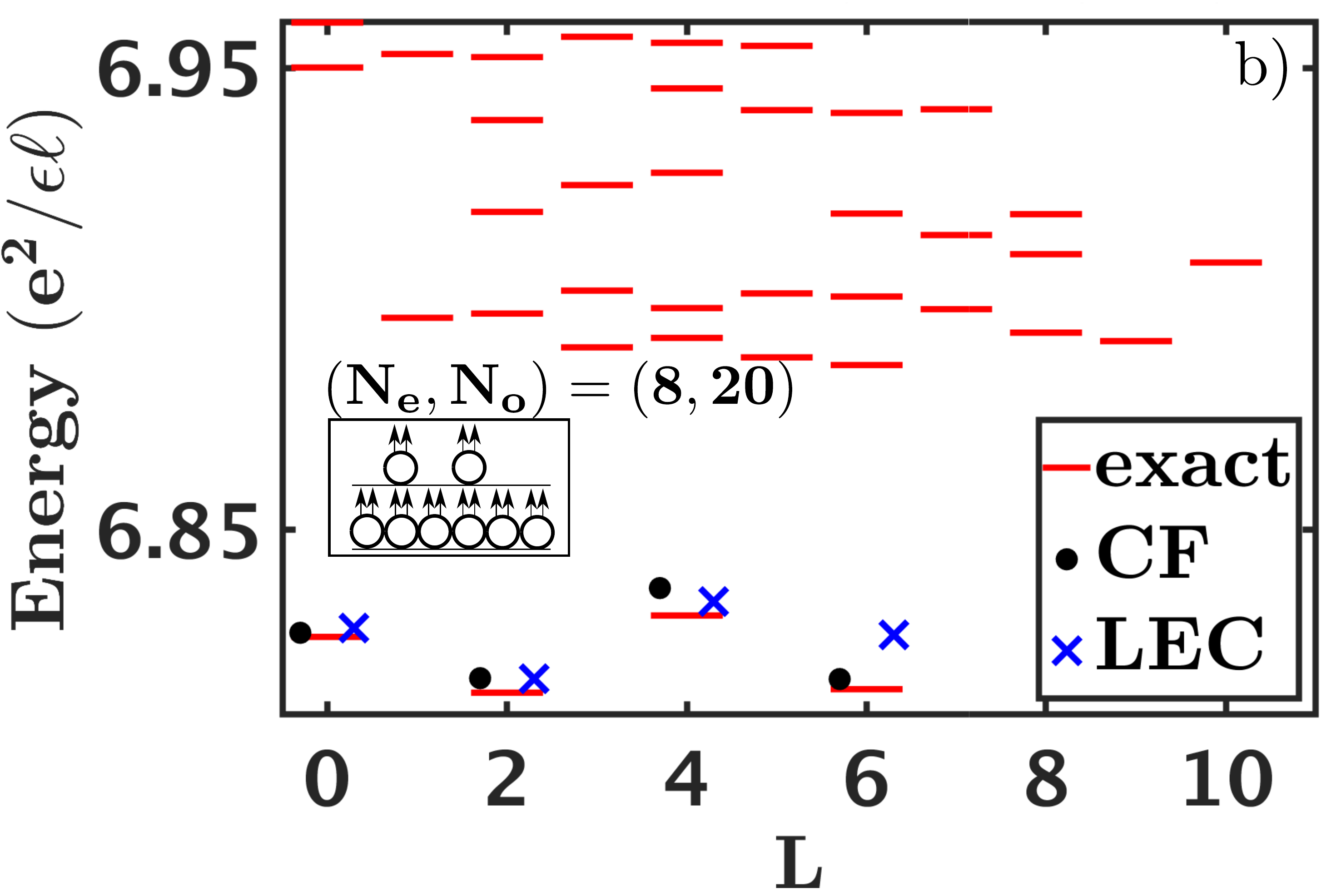} 
\includegraphics[width=0.32\textwidth]{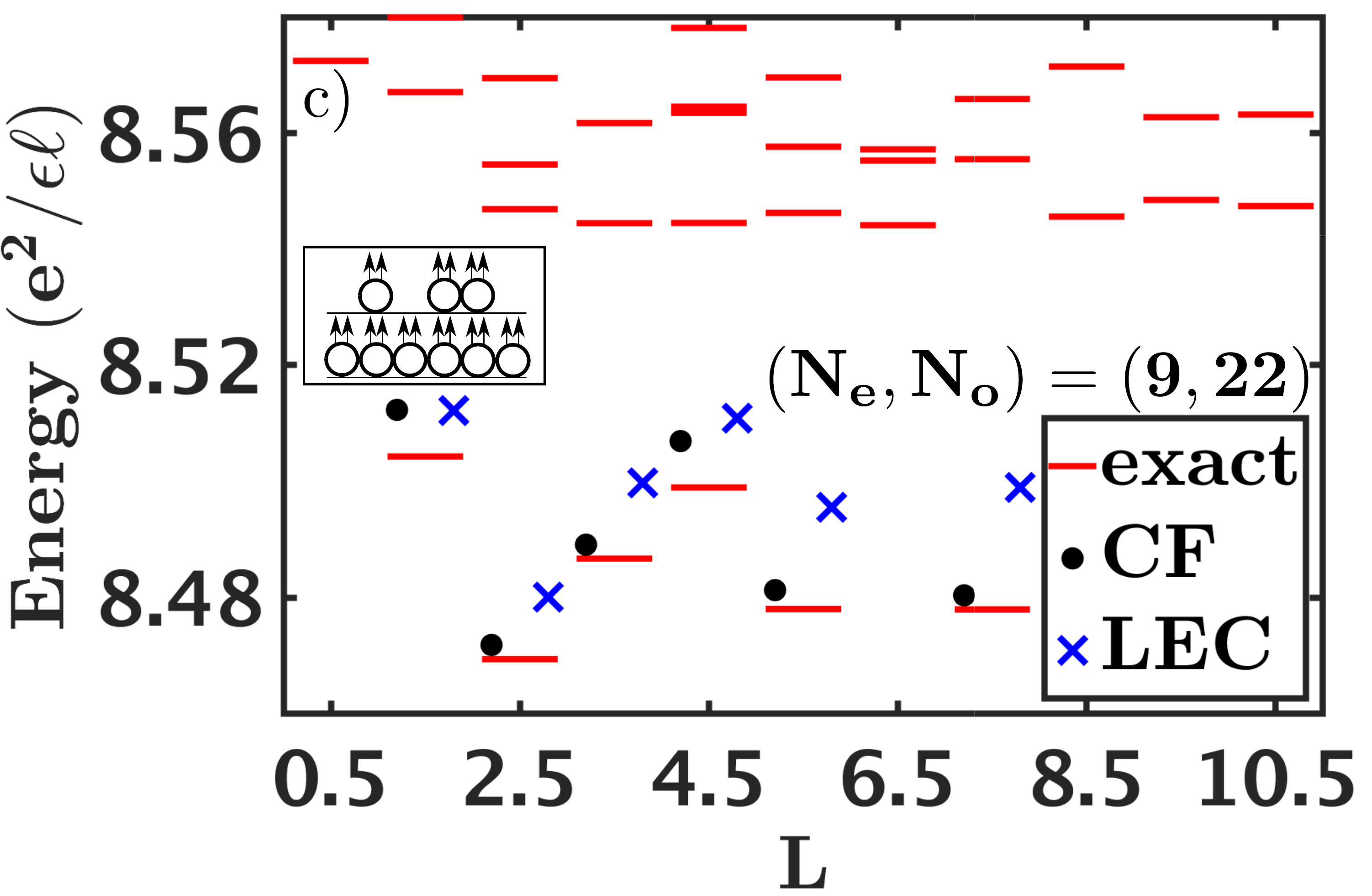} \\
\includegraphics[width=0.32\textwidth]{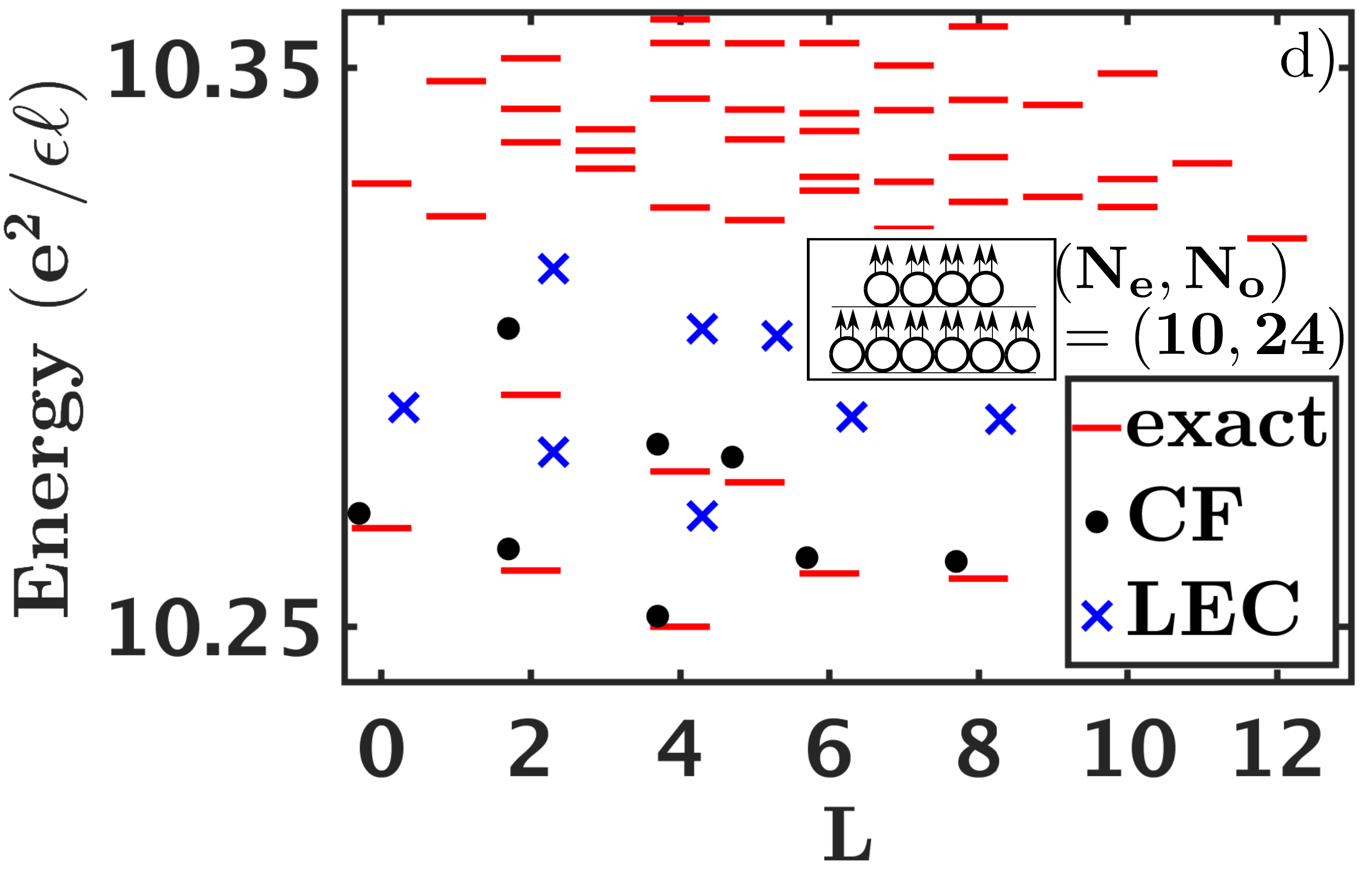}
\includegraphics[width=0.32\textwidth]{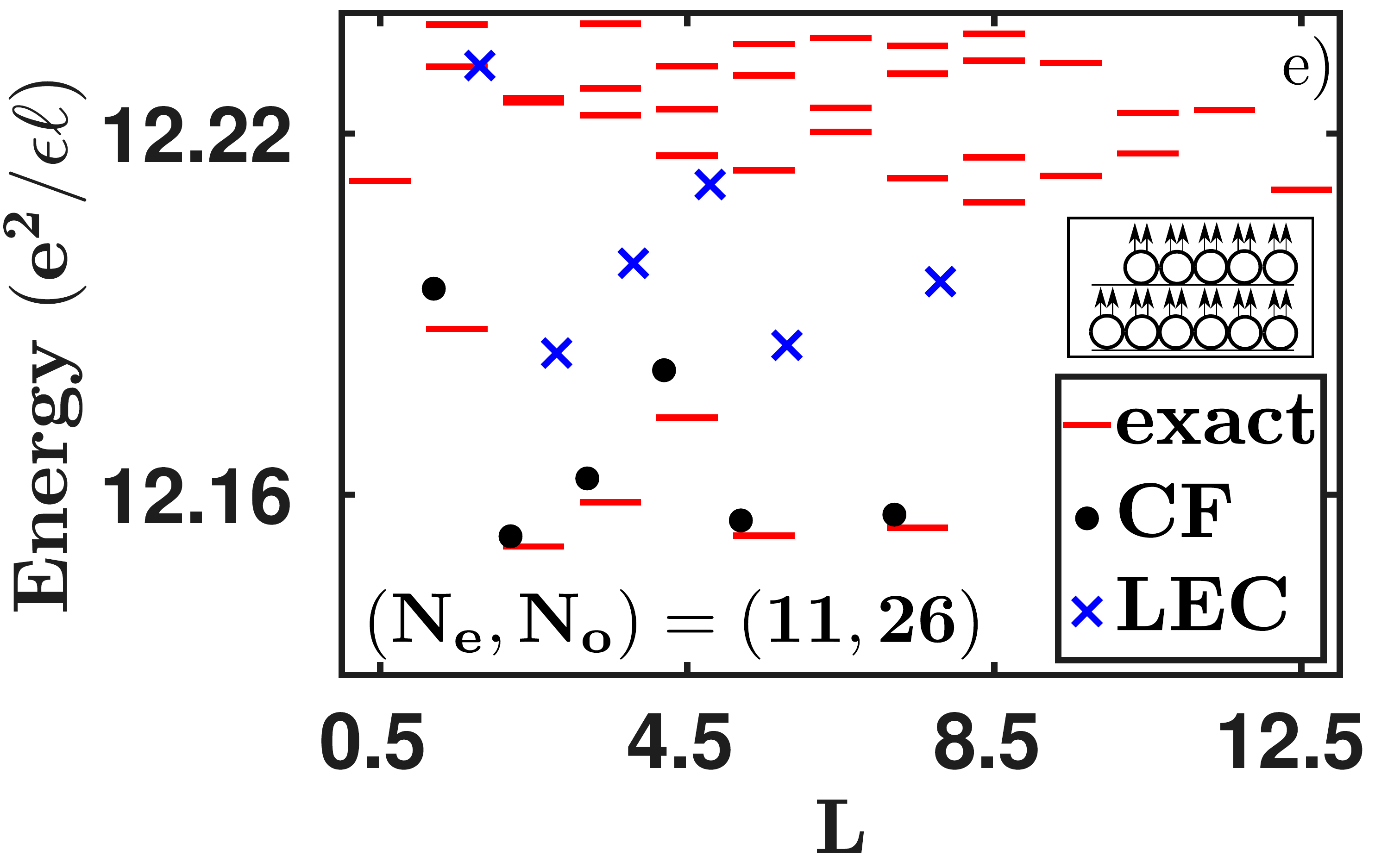}
\caption{Comparison of the lowest Landau level Coulomb spectra obtained from exact diagonalization (red dashes), composite fermion theory (black dots) and local exclusion conditions (blue crosses) for systems consisting of one (a), two (b), three (c), four (d) and five (e) quasielectrons at $\nu=1/3$. The calculations were carried out in the spherical geometry with $N_{e}$ electrons in $N_{o}$ orbitals. The composite fermion states are shown schematically in the inset.}
\label{fig:LEC_vs_CF_vs_exact}
\end{figure*}

The wave functions of Eq.~(\ref{eq_Jain_CF_wf}) are most readily evaluated in first quantization using the Metropolis Monte Carlo method~\cite{Binder10}. In Fig.~\ref{fig:LEC_vs_CF_vs_exact} we show the LLL Coulomb spectra of multiple quasielectron states at $\nu=1/3$ obtained from the LEC and CF constructions. We first point out that the angular momentum quantum number for a state with a single quasielectron obtained from the LEC and CF constructions agree with each other. Moreover, their LLL Coulomb energies are close to each other, indicating high overlaps of the corresponding wave functions. Furthermore, the overlaps and energies of multiple quasielectron states constructed from the LEC and CF theories are also in fairly good agreement with each other. The CF wavefunctions are known to be very accurate for the LLL Coulomb interaction and thus, in general, have lower variational energies than the LEC states. The LEC formalism only defines the universal sub-Hilbert spaces of quasielectrons of different topological phases and is thus not specifically optimized for a particular interaction. The LLL Coulomb interaction leads to scattering between different types of quasielectrons, which is ignored in obtaining the variational energies of the LEC states in Fig.(\ref{fig:LEC_vs_CF_vs_exact}). This leads to higher variational energies in certain cases. We expect shorter-range interactions to reduce scattering between different quasielectron types, thereby lowering the variational energies of the LEC states. Physically this implies the topological properties of the quasielectrons are more robust with such short-range interactions, as compared to the LLL Coulomb one.

It is important to note that the Coulomb interaction, unlike the short-range model $V_{1}$ interaction, is long-ranged. Therefore, a reasonable agreement in the LLL Coulomb energies of the CF and LEC states suggests that the two seemingly disparate constructions are consistent with each other. Finally, we mention that both the LEC and CF constructions give a good representation of the LLL Coulomb spectra as evidenced from their comparison with the spectra obtained from exact diagonalization [see Fig.~\ref{fig:LEC_vs_CF_vs_exact}]. The qualitative and semi-quantitative agreement between the LEC and the CF construction for the Laughlin quasielectrons has important implications. In particular, the Gaffnian and the CF states at $\nu=2/5$ turn out to be closely related. This is because the former is made of undressed $\left(5,2\right)\text{Type}$ quasielectrons, while the latter is made of undressed CF quasielectrons. Our analysis in this section shows that for undressed quasielectrons, the LEC and CF constructions produce physically equivalent states. When quasielectrons are dressed with neutral excitations, however, the LEC and the CF constructions have qualitatively different predictions in terms of the counting of the states. Typically, the manifold of dressed CF quasielectrons is much larger than the manifold of dressed LEC quasielectrons of the same type. A detailed study of these connections will be presented elsewhere.

\section{Conclusions}\label{summary}
In this work, we extended the recently developed LEC construction~\cite{Yang19} to study the elementary excitations, namely quasiholes, quasielectrons, and neutral excitations, of many FQH phases. The quasihole model wave functions can be generated using the same set of LECs that determine the corresponding ground state. This can be achieved by imposing the highest weight condition on the truncated Hilbert space containing more orbitals than the ground state (with the same number of electrons as in the ground state). Empirically, we find the edge modes of the FQH phase derived from the quasihole counting matches exactly with the counting of the entanglement spectrum obtained from the ground state. Such agreement is known to hold for the FQH states constructed from a conformal field theory (CFT). Our studies indicate that such a bulk-edge correspondence holds more generally, even for states that do not lend themselves to a CFT description. 

In contrast to the quasiholes, the set of LECs defining the quasielectrons is different from but closely related to, the set that identifies the ground state. This is reasonable since unlike quasiholes, quasielectrons cost finite energy in the thermodynamic limit which arises from violating the commensurability conditions of the FQH ground state. It is well-known that an FQH phase can have different types of quasielectrons. Here we showed that the quasielectron manifold, which is composed of different types of quasielectrons, can be uniquely determined by the LEC construction. The neutral excitations can be studied analogously as bound states of quasielectrons and quasiholes. The LEC scheme thus allows us to systematically build the entire energy spectrum of an FQH phase with different types of elementary excitations. In particular, the many-body wave functions for the quasielectrons and neutral excitations can be explicitly constructed and uniquely determined once the LEC conditions are specified. However, it is not easy to generically write these wave functions down in a simple first quantized form, expect for a few special cases~\cite{Yang13a}.

We also identified several cases where a set of LECs defining the quasielectrons of one FQH phase also defined the ground state and quasiholes of a different FQH phase. In particular, we explicitly showed that the Gaffnian ground state at $\nu=2/5$ is made of a particular type of Laughlin quasielectrons. For undressed quasielectrons at $\nu=1/3$, we provided \emph{numerical evidence} to show that the LEC and the standard CF constructions, which are microscopically different, lead to topologically equivalent states. Given that the Abelian Jain $\nu=2/5$ state is made of CF quasielectrons, we conjecture that just from the wave function itself, the Gaffnian ground state and the Jain $\nu=2/5$ state are physically indistinguishable. All topological indices that can be extracted from the two wave functions could be identical. Furthermore, the Fibonacci ground state at $\nu=3/5$ is made of a particular kind of Moore-Read quasielectrons. Thus, the LEC construction opens an avenue to find novel connections between various FQH states.

The LEC construction shows that, besides the ground state, the excitations of FQH fluids are also determined by the algebraic structure of the truncated Hilbert space. This is because the quasihole and quasielectron manifold (and thus the neutral excitation manifold) can be defined without referring to microscopic Hamiltonians. Indeed, the topological properties of a particular FQH phase characterize both the ground state as well as the excitations. For the ground state, the topological indices include the filling factor, the topological shifts, the topological entanglement entropy, and particle clustering. For quasiholes and quasielectrons, the topological indices include their fractionalized charge, the quasihole counting, and the topological spins. We have shown here that the LEC construction can uniquely determine all these topological indices, and thus the universal properties of many FQH phases. It would be interesting to further explore how robust these universal topological properties are in the presence of realistic interactions. The study of the interplay between the Hilbert space algebra and a realistic interaction could have crucial experimental ramifications, especially for non-Abelian FQH phases.

\begin{acknowledgments}
We thank Jainendra Jain, Ying-Hai Wu, and Zlatko Papic for useful discussions. This work is supported by the NTU grant for Nanyang Assistant Professorship. The Center for Quantum Devices is funded by the Danish National Research Foundation. This work was supported by the European Research Council (ERC) under the European Union Horizon 2020 Research and Innovation Programme, Grant Agreement No. 678862 and the Villum Foundation. Some of the numerical calculations reported in this work were carried out on the Nandadevi supercomputer, which is maintained and supported by the Institute of Mathematical Science’s High Performance Computing center.
\end{acknowledgments}

\bibliography{biblio_fqhe}
\end{document}